\documentclass[journal]{IEEEtran}
\usepackage{epsf}
\newcommand{\ul}[1]{\underline{#1}}

\newcommand{\bea}{\begin{eqnarray}}

\newcommand{\eea}{\end{eqnarray}}

\newcommand{\rmc}{{\rm c}}

\newcommand{\rmd}{{\rm d}}

\newcommand{\rmj}{{\rm j}}

\newcommand{\rmr}{{\rm r}}

\newcommand{\rmw}{{\rm w}}

\newcommand{\rmu}{{\rm u}}

\newcommand{\Beta}{{\rm B}}

\newcommand{\med}{{\rm med}}

\newcommand{\mod}{{\rm mod}}

\begin{document}


\title{Probability Distribution of the Quality Factor of a Mode-Stirred Reverberation Chamber}

\author{
	Luk R. Arnaut and Gabriele Gradoni
\\ 
Department of Electrical and Electronic Engineering\\ Imperial College of Science, Technology and Medicine\\London SW7 2AZ, United Kingdom\\ (l.arnaut@imperial.ac.uk)\\
and\\
Institute for Research in Electronics and Applied Physics\\
University of Maryland\\
College Park
20742 MD, United States\\
(ggradoni@umd.edu)
}
\date{\today}

\maketitle

\begin{abstract}
We derive a probability distribution, confidence intervals and statistics of the quality ($Q$) factor of an arbitrarily shaped mode-stirred reverberation chamber, based on ensemble distributions of the idealized random cavity field with assumed perfect stir efficiency. It is shown that $Q$ exhibits a Fisher--Snedecor F-distribution whose degrees of freedom are governed by the number of simultaneously excited cavity modes per stir state. The most probable value of $Q$ is between a fraction $2/9$ and $1$ of its mean value, and between a fraction $4/9$ and $1$ of its asymptotic (composite $Q$) value. The arithmetic mean value is found to always exceed the values of all other theoretical metrics for centrality of $Q$. For a rectangular cavity, we retrieve the known asymptotic $Q$ in the limit of highly overmoded regime. 
\end{abstract}

\section{Introduction \label{sec:intro}}
To date, the study of mode-tuned or mode-stirred reverberation chambers (MT/MSRCs) 
 -- i.e., multi-mode cavity resonators furnished with a 'stirring' mechanism that produces time-varying excitation and/or boundary conditions -- has mainly focused on the properties of the random electromagnetic (EM) {\em field}. 
Probability density functions (PDFs) for idealized and imperfect fields, including EM boundary-value problems \cite{dunn1}, \cite{arna2007}, were calculated and compared with measurements or simulations.
A natural extension is the stochastic characterization of {\em intrinsic EM parameters\/} of instrumentation and devices subjected to random fields, e.g., wave and input impedances \cite{water1963}--\cite{zheng2006}, antenna parameters {\cite{warn1},} \cite{junq1}, etc.

One of the fundamental parameters of a MT/MSRC is its quality ($Q$-) factor \cite{junq1}--\cite{iec}.
In the simplest model, $Q$ is defined by a constant single value, as the ratio of the stir-averaged\footnote{The fact that the total energies stored and dissipated vary with changing stir state has been validated by experiments, which show that the measurement of $S_{11}$ at a fixed frequency and source power $P_{s}$ exhibits random fluctuations with changing stir state. Therefore, the net forward power $P_{\rm Tx} = (1-|S_{11}|^2)P_{s}$ injected into the MT/MSRC fluctuates accordingly.} 
stored energy $\langle U \rangle$ to the averaged dissipated power $\langle P_\rmd \rangle$, multiplied by the excitation frequency\footnote{{In general, the spectral power density $g(\omega)$ and, hence, the spectrally averaged angular centre frequency $\omega_0 = \int^{\infty}_0 \omega g(\omega) \rmd \omega / \int^{\infty}_0 g(\omega) \rmd \omega$ vary as a function of stir state \cite{arnalocavg}. Therefore, spectral and ensemble averagings are strictly needed to replace $\omega$ in (\ref{eq:defQeff}) by $\langle \omega_0 \rangle$. For narrowband excitation or nondispersive $g(\omega)$, the fluctuations of $\omega_0$ are usually negligibly small, whence ensemble and spectral averaging of $\omega$ can then be omitted.}} $\omega$ \cite{lamb1}--\cite{liu1}:
\bea
Q_{\rm eff}(\omega) \stackrel{\Delta}{=} \omega \frac{\langle U (\omega) \rangle}{\langle P_\rmd (\omega) \rangle} \label{eq:defQeff}
.
\eea 
The definition of this so-called `effective' or `composite' quality factor is inspired by the corresponding notion of modal $Q$ for a single eigenmode of a static resonant cavity, i.e., 
\bea
Q_{mnp} (\omega_{mnp}) \stackrel{\Delta}{=} \omega_{mnp} \frac{U_{mnp}(\omega_{mnp})}{P_{\rmd,mnp}(\omega_{mnp})}, \label{eq:defQmodal}
\eea
in which $\omega_{mnp}$, $U_{mnp}$ and $P_{\rmd, mnp}$ take constant values for a selected mode specified by modal indices $m$, $n$, $p$. 
In a MT/MSRC, however, $U$ and $P_\rmd$ fluctuate quasi-randomly as a function of stir state $\tau$. 
Hence, defining an instantaneous value of $Q$ at each $\tau$ as
\bea
Q(\omega,\tau) \stackrel{\Delta}{=} \omega(\tau) \frac{U(\omega, \tau)}{P_{{\rm d}}(\omega, \tau)}, \label{eq:defQ}
\eea 
this $Q$ is now a {\em randomly fluctuating\/} quantity with an associated PDF $f_Q(q)$, correlation functions, etc., 
when considered across all $\tau$. 
For simplicity of notation, we shall further omit indicating the dependencies on $\omega$ and $\tau$ in (\ref{eq:defQ}).

Compared to (\ref{eq:defQeff}), the definition (\ref{eq:defQ}) is closer in spirit to the original concept of $Q_{mnp}$. Firstly, (\ref{eq:defQ}) involves a ratio of quantities that exist physically at each $\tau$, as opposed to the formal ratio of mean values in (\ref{eq:defQeff}) that exist only in a $\tau$-averaged, i.e., mathematical sense. Secondly, because of propagation of uncertainties, any disregard for the random fluctuations of $U$ and $P_\rmd$ results in an underestimate of the level of fluctuation of other stochastic EM quantities that depend explicitly or implicitly on $Q$, in {particular} the standard deviations of the EM fields $\sigma_{E^{\prime(\prime)}_\alpha}$ and $\sigma_{H^{\prime(\prime)}_\alpha}$ (cf. (\ref{eq:sigmaHsigmaE})), which are of fundamental importance.

An alternative but more restrictive approach to quantifying the uncertainty of $Q$ was developed previously in \cite{arnaQ}. There, second-order statistical characterization of $Q$ was performed based on spectral and ensemble averaging of $Q_{mnp}$ for TE and TM eigenmodes. This permitted a calculation of the mean $\mu_Q$ and standard deviation $\sigma_Q$ for a rectangular cavity in which wall stirring causes modal perturbations. 

In the present paper, the use of sampled instantaneous (as opposed to averaged) values of $Q$ allows for the calculation of the complete PDF $f_Q(q)$. This provides a more comprehensive characterization compared to mere first- and second-order moments. 
For simplicity, the analysis is based on {\em ensemble\/} distributions of $U$ and $P_{\rmd}$ for ideal Gaussian EM fields, as opposed to their sampling distributions \cite{arnaTQE2}, \cite{arnaPRE}. This implicitly assumes that a sufficiently large and theoretically infinite number ($N$) of statistically independent stir states for the field is generated by the stir process ($N\rightarrow+\infty$).
It will be found that $f_Q(q)$ then satisfies a Fisher--Snedecor F-distribution, whose two numbers of degrees of freedom (DoF) both depend on the number ($M$) of simultaneously excited cavity modes per stir state. When $M$ increases, $f_Q(q)$ evolves from a positively skewed PDF for low $M$ toward a Gaussian (normal) PDF, accompanied by a reduction in mean value and absolute or relative spread of $Q$.

The results apply generally to cavities with arbitrary geometries, including irregular shapes, but will be illustrated with explicit expressions for simple (integrable) rectangular cavities.
Except in Sec. \ref{sec:MSvsMT}, we do not distinguish between mode-tuned and mode-stirred methods of operation, insofar as only quasi-static
fields are considered. An $\exp(\rmj \omega t)$ time dependence is assumed and suppressed throughout. 
Different types of averaging will be performed: we shall use the notations $\langle \cdot \rangle_V$, $\langle \cdot \rangle_S$ and $\langle \cdot \rangle$ to represent spatial averaging with respect to the volume $V$, surface area $S$, and ensemble averaging with respect to cavity stir states $\tau$, respectively. The ensemble average assumes equal $V$ and $S$ throughout, as a prerequisite for constant average spectral mode density, and only involves perturbations of shape or aspect ratio(s) across different realizations.

\section{Unstirred chambers with single-mode excitation: deterministic $U$ and $P_{\rmd}$\label{sec:unstirred}}
For the purpose of establishing definitions and extending the analysis later to stirred multi-mode cavities, we briefly review some basic results for the $Q$ of a single eigenmode in a static (unstirred) cavity at its fixed angular resonance frequency $\omega_{mnp}$. In this case, the modal $Q_{mnp}$ is single-valued.
In an unstirred cavity, the local electric and magnetic modal amplitudes $E_0$ and $H_0$ at any location $\ul{r}$ inside $V$ are time invariant.  
For a lossless linear time-invariant isotropic homogeneous medium filling the cavity interior, $\ul{D}=\epsilon_0\ul{E}$ and $\ul{B}=\mu_0\ul{H}$, whence the electric and magnetic stored energies $U_{\rm e}$ and $U_{\rm m}$ are proportional to the spatial integrals of the local intensities $|\ul{E}(\ul{r})|^2$ and $|\ul{H}(\ul{r})|^2$, respectively. 
For steady state excitation, $U_{\rm e}=U_{\rm m}$ to first approximation at sufficiently high frequencies. 
The total stored energy $U=U_{\rm e}+U_{\rm m}$ can thus be expressed as
\bea
U
&=&\frac{1}{2} \int_V \left [ \ul{E}(\ul{r}) \cdot \frac{\partial \ul{D}^*(\ul{r})}{\partial t}+ \ul{H}(\ul{r}) \cdot \frac{\partial \ul{B}^*(\ul{r})}{\partial t} \right ] {\rm d}V \nonumber\\
&=&\frac{1}{2} \int_V \left [ \frac{\epsilon_0}{2} |\ul{E}(\ul{r})|^2 + \frac{\mu_0}{2} |\ul{H}(\ul{r})|^2 \right ] {\rm d}V \nonumber\\
&=& \frac{\mu_0}{2} \int_V |\ul{H}(\ul{r})|^2 {\rm d}V 
\stackrel{\Delta}{=} \frac{\mu_0 \thinspace V}{2} \langle |\ul{H}|^2 \rangle_V,
\label{eq:U}
\eea 
where an asterisk denotes complex conjugation. 

To obtain a corresponding expression for $P_\rmd$, the relevant quantity is the tangential magnetic field $\ul{H}_t$ at a location $\ul{r}_S$  on the cavity's interior boundary surface $S=\partial V$ with unit local inward surface normal $\ul{1}_n(\ul{r}_S)$.  
Conduction loss in the wall yields a nonvanishing tangential electric field $\ul{E}_t = R_\rmw ( \ul{H}_t \times \ul{1}_n ) \not = \ul{0}$ at $S$ that can be envisaged as a surface layer of magnetic current produced by $\ul{H}_t$ as an equivalent boundary source and dissipated by $S$, in addition to the sheet of surface charge produced by the normal electric field in a lossless cavity.
The time-averaged absorbed energy is the spatially integrated real part of the normal component of the local Poynting vector, ${S}_n={\rm Re}[\ul{1}_n \cdot ( \ul{E}\times \ul{H}^*) ]/2$, i.e., 
\bea
P_{\rm d} &=& \frac{R_\rmw}{2} \int_S \left | \ul{1}_n (\ul{r}_S) \times \ul{H}(\ul{r}_S) \right |^2 \rmd S \nonumber\\
&=& \frac{1}{2\sigma_\rmw \delta_\rmw} \int_S \left | \ul{H}_t(\ul{r}_S) \right |^2 \rmd S 
 \stackrel{\Delta}{=} \frac{S}{2\sigma_\rmw \delta_\rmw} \langle \left | \ul{H}_t \right |^2 \rangle_S ,
\label{eq:Pd}
\eea
in which 
$R_\rmw \stackrel{\Delta}{=} 1/(\sigma_\rmw \delta_\rmw) = \sqrt{\omega \mu_\rmw/(2\sigma_\rmw)}$ represents the per-unit area surface resistance of the interior cavity wall, $\delta_\rmw \stackrel{\Delta}{=} \sqrt{2/(\omega \mu_\rmw \sigma_\rmw)}$ is its skin depth, and where $\mu_\rmw=\mu_{\rmw,\rmr} \mu_0$ and $\sigma_\rmw$ are its permeability and conductivity, respectively. Substituting (\ref{eq:U}) and (\ref{eq:Pd}) into (\ref{eq:defQ}) yields \cite[sec. 10.4]{vanb2}
\bea
Q =
\frac{2}{\mu_{\rmw,\rmr} \delta_\rmw} \frac{\int_V |\ul{H}(\ul{r})|^2 \rmd V}{\int_S |\ul{H}_t(\ul{r}_S)|^2 \rmd S}
=
\frac{2\thinspace V}{\mu_{\rmw,\rmr} \delta_\rmw\thinspace S} \frac{\langle |\ul{H}|^2 \rangle_V}{\langle |\ul{H}_t|^2 \rangle_S}
.
\label{eq:Qgenfinal}
\eea
If only one resonant mode is excited, then the local field amplitudes throughout $V$ are characterized by a single modal amplitude value $H_{mnp,0}$, whence (\ref{eq:Qgenfinal}) can {then} be written as
\bea
Q =
\frac{2}{\mu_{\rmw,\rmr} \delta_\rmw} \frac{|H_{mnp,0}|^2}{|H_{mnp,t,0}|^2 } \frac{\int_V |\ul{\phi}_{mnp}(\ul{r})|^2 \rmd V}{\int_S |\ul{\phi}_{mnp}(\ul{r}_S)|^2 \rmd S},
\eea
where $\ul{\phi}_{mnp}$ is the real-valued magnetic eigenvector, and $H_{mnp,(t,)0}$ is its associated complex-valued amplitude.

As an example, consider a rectangular cavity with $V=\ell_x \ell_y \ell_z$ with a single excited mode whose local amplitude $|\ul{H}_{mnp}(\ul{r})|$ at $\ul{r}=x\ul{1}_x + y\ul{1}_y + z\ul{1}_z$ can be expressed as
\bea
|\ul{H}_{mnp}(\ul{r})|&=& |{H}_{mnp,0}| 
\left \{ \begin{array}{l} \sin(k_{mnp,x} x)\\ \cos(k_{mnp,x} x) \end{array} \right \} \nonumber\\
&~& \cdot
\left \{ \begin{array}{l} \cos(k_{mnp,y} y)\\ \sin(k_{mnp,y} y) \end{array} \right \} \cdot
\left \{ \begin{array}{l} \sin(k_{mnp,z} z)\\ \cos(k_{mnp,z} z) \end{array} \right \}.
\label{eq:amplit}
\eea
A similar expression for the electric field $\ul{E}_{mnp}(\ul{r})$ with amplitude
$|E_{mnp,0}|$ applies. 
Thus, the amplitudes are unmodulated (sinusoidal or constant) with respect to $\ul{r}$, in all directions.
For a transverse mode, any valid combination in (\ref{eq:amplit}) consists of two {spatial} harmonic functions along two orthogonal directions $\ul{1}_\alpha$ and $\ul{1}_\beta$ combined with the unit function in the third direction $\ul{1}_\gamma$ (i.e., $\cos(k_{mnp,\gamma} \gamma)=1$), where $\alpha,\beta,\gamma \in \{x,y,z\}$.
However, in MT/MSRCs, we are primarily interested in overmoded conditions at high frequencies, where the vast majority of modes are hybrid.
For a hybrid mode satisfying the EM boundary conditions, a valid combination in (\ref{eq:amplit}) is the product of three such harmonic functions.
Substituting (\ref{eq:amplit}) into (\ref{eq:U}) and (\ref{eq:Pd}), {together} with $\int^{\ell_\alpha}_0 \sin^2(k_{mnp,\alpha} \alpha) \rmd \alpha = \int^{\ell_\alpha}_0 \cos^2(k_{mnp,\alpha} \alpha) \rmd \alpha = \ell_\alpha/2$, we obtain for a rectangular cavity with conducting boundaries
\bea
U = \frac{\mu_0 \thinspace V}{16} |H_{mnp,0}|^2,~~P_\rmd = \frac{S}{8} \sqrt{\frac{\omega\mu_\rmw}{2\sigma_\rmw}} |H_{mnp,t,0}|^2,
\label{eq:UPdsinglemodeunstirred}
\eea
in which the difference between the factors $1/16$ and $1/8$ results from the fact that
a hybrid mode generally exhibits three magnetic field components for the interior field, whereas only two nonvanishing tangential components of this field exist on the surface.
For a transverse mode, the factors $1/16$ and $1/8$ in (\ref{eq:UPdsinglemodeunstirred}) are replaced by $1/8$ and $1/4$ (or possibly $1/2$, but with vanishingly small contribution, when the cavity surface is locally perpendicular to the transverse direction of the mode), respectively.
From (\ref{eq:defQ}) and (\ref{eq:UPdsinglemodeunstirred}), it follows that
\bea
Q = \frac{V}{\mu_{\rmw,\rmr} \delta_\rmw S} \frac{|{H}_{mnp,0}|^2}{|{H}_{mnp,t,0}|^2}
.
\label{eq:Qsinglemodeunstirred}
\eea

For any single mode in a rectangular unstirred cavity, $H_{mnp,0}$ and $H_{mnp,t,0}$ are constant with respect to location and time.
In nonrectangular (e.g., cylindrical) cavities, the eigenmodes no longer consist of spatial harmonics. For nonseparable geometries, they may not even exist in closed-form expressions. Hence the ratio $\langle |\ul{H}(\ul{r})|^2 \rangle_V / \langle |\ul{H}_t(\ul{r}_S)|^2 \rangle_S$ is in general different from $|{H}_{mnp,0}(\ul{r})|^2/(2|{H}_{mnp,t,0}(\ul{r}_S)|^2)$, in which case (\ref{eq:Qsinglemodeunstirred}) is then multiplied by a shape dependent factor $h$ (cf. (\ref{eq:defh})).

\section{Stirred chambers: quasi-random $U$ and $P_{\rmd}$}
\subsection{Single vs. multimode excitation \label{sec:singlevsmulti}}
When invoking mode tuning or mode stirring, the eigenmodes and therefore $Q$ evolve with stir state $\tau$. The characterization in Sec. \ref{sec:unstirred} for a single mode is then only meaningful in a statistical sense. We denote $\{ \ul{E}(\tau|\ul{r}) \}$ and $\{ \ul{H}_{(t)}(\tau|\ul{r}_{(S)}) \}$ to represent ensembles of sample sets of $N$ stir states of the field at arbitrary $\ul{r}_{(S)}$. 
These stirred local fields vary randomly in spatial orientation, magnitude and phase as a function of $\tau$. Generally, $\ul{H}(\tau|\ul{r})$ and $\ul{H}_{t}(\tau|\ul{r}_{S})$ have three and two nonzero complex-valued (in-phase and quadrature) rectangular components $H_\alpha$ ($\alpha=x,y,z$), respectively. Hence, if at any $\tau$ only one cavity mode is excited that is randomly perturbed by the stir process ({i.e.,} random single-mode excitation), then the local fields $\ul{H}(\tau|\ul{r})$ and $\ul{H}_{t}(\tau|\ul{r}_S)$ in (\ref{eq:U})--(\ref{eq:Qgenfinal}) are now {\em random\/} processes with six and four DoF, respectively. 

In practical (non-superconducting) overmoded MT/MSRCs, significant multimode excitation occurs typically. Spectral overlap of nondegenerate modes having nonzero absorption bandwidths causes intermodal coupling, such that even a single-frequency (CW) source then excites simultaneously several modes with different $\omega_{mnp}$.
Alternatively, the spectrum of a wide-band source may encompass several $\omega_{mnp}$ of the MT/MSRC, so that corresponding (non)overlapping modes may be simultaneously excited (e.g., multitones in a multimode laser or in certain communications protocols). 
Let the number of simultaneously excited modes per stir state be denoted by $M$. 
A physical estimation of the value of $M$ is given in Appendix \ref{app:estM}. 
For arbitrary $\tau$, the cavity field is the resultant of the weighted superposition of $M$ participating modes (random walk model), producing a spatial modulation of the amplitudes $E_{0}$ and $H_{(t,)0}$ across the cavity's interior.
If the structure of the cavity is sufficiently irregular at $\omega$, then this spatial distribution is quasi-random.
This (static) spatial variation is additional to the (dynamic) fluctuations of the local field caused by stirring and yields a 4-D spatio-temporal random field.
For spatially random fields, (\ref{eq:U}) and (\ref{eq:Pd}) remain valid in a statistical sense, i.e., at arbitrary $\tau$.
Thus, for a wall-stirred rectangular cavity, we now have instead of (\ref{eq:UPdsinglemodeunstirred})--(\ref{eq:Qsinglemodeunstirred}), 
\bea
U(\tau) &=& \frac{\mu_0 \thinspace V}{16} \langle |{H}_0(\tau)|^2 \rangle_V,\label{eq:U_ifv_tau}\\
P_\rmd(\tau) &=& \frac{S}{8} \sqrt{\frac{\omega\mu_\rmw}{2\sigma_\rmw}} \langle |{H}_{t,0}(\tau)|^2 \rangle_S,\label{eq:Pd_ifv_tau}\\
Q(\tau) &=& \frac{V}{\mu_{\rmw,\rmr} \delta_\rmw S} \frac{\langle |H_{0}(\tau)|^2 \rangle_V}{\langle |H_{t,0}(\tau)|^2 \rangle_S}.
\label{eq:UPdmultimodestirred}
\eea

Each individual mode acts as a `channel' for storage and dissipation of energy and increases by one unit the number of ways in which the value of each Cartesian component of the resultant field phasor can be obtained. The DoFs of the spatially integrated field intensities leading to $U$ and $P_\rmd$ increase accordingly.
For each mode, $\ul{E}_{mnp}$ and $\ul{H}_{mnp}$ are physically (i.e., deterministically) related via a wave impedance dyadic. Following (\ref{eq:U})--(\ref{eq:Pd}), the increase of the number of DoF of $U$ by each mode is therefore the same as for $U_{\rm e}$ and $U_{\rm m}$ individually, i.e., six, whereas the corresponding increase for $P_\rmd$ is four. 

Finally, 
assuming that the stirring process is sufficiently efficient to be capable of generating a very large (theoretically infinite) value of $N$ that produce independent and identically distributed ideal Gaussian $\ul{H}_{(t)}$ (i.e., $N\rightarrow +\infty$)
and 
assuming that wide-sense ergodicity of the fields holds (i.e., 
$\mu_{\ul{H}(\ul{r}|\tau)} = \mu_{\ul{H}(\tau|\ul{r})}$ and 
$\sigma_{\ul{H}(\ul{r}|\tau)} = \sigma_{\ul{H}(\tau|\ul{r})}$), 
such that the spatial distributions of each ${H}_{(t,)\alpha}$ is identical to its ensemble (i.e., stir) distribution,
$U$ then  exhibits approximately\footnote{The incoherent superposition of $M$ participating modes with equal $\chi^2_6$ energy distributions presumes that energy is equally partitioned across these modes. For overlapping modes, this is only approximately true because the partitioning depends on the source's spatial location and on the specific spectral distances of the $\omega_{m^\prime n^\prime p^\prime}$ relative to the excitation frequency $\omega$. They should be Lorentz weighted accordingly in the superposition.} 
a $\chi^2_{6M}$ PDF, and $P_{\rmd}$ has approximately\footnote{On $S$, the nonorthogonality of irrotational and solenoidal magnetic eigenvectors (i.e., solutions associated with boundary conditions of magnetic type) causes the total power loss to deviate from the sum of the power losses of individual modes \cite[Sec. 10.4]{vanb2}. This nonorthogonality results in the DoF $2s$ to be somewhat less than $4M$.} 
a $\chi^2_{4M}$ PDF across the stir states and cavity interior, i.e., (\ref{eq:PDFU}) and (\ref{eq:PDFPd}) hold with $r=3M$ and $s=2M$, respectively.  

In practice, the values of $M$ and $N$ evaluated for different stir processes are often strongly positively correlated. For small $M$, the practical stirring performance may then be significantly compromized. If the assumption $N\rightarrow +\infty$ becomes unsustainable, then the use of Bessel $K$ sampling distributions for $\ul{E}$, $\ul{H}_{(t)}$, $U$ and $P_\rmd|U$ \cite{arnaPRE} offers an appropriate framework for characterizing $f_Q(q)$ \cite{arnaESA}. 
In practice, the case $M=1$ often (although not exclusively) involves excitation at wavelengths that are not small compared to cavity dimensions, while also the modal overlap is small. In this case, the stir process is also typically (but not inevitably) less efficient, whence the $\chi^2_{6}$ and $\chi^2_{4}$ PDFs of $U(\tau|\ul{r})$ and { $P_\rmd(\tau|\ul{r}_S)|U(\tau|\ul{r})$} and, hence, the PDF (\ref{eq:PDFQ_selfsuff_mu_multimod}) are then only approximately valid. 

Although we shall further focus on the case where $M$ is a constant integer with respect to stir state, one may envisage a situation where its value could fluctuate as a function of $\tau$. In this case, we can estimate $M$ by its mean value $\sum^N_{\tau=1} M(\tau)/N$, which may be fractional, yielding a generalization of $\chi^2_{4M}$ and $\chi^2_{6M}$ PDFs to gamma PDFs. Values of $M$ smaller than unity represent the case where, on average, less than one mode per stir state is being excited.

\subsection{Probability density function and statistics of $Q$ \label{sec:PDFQ}}

\begin{figure}[htb] \begin{center} \begin{tabular}{l}
\ \epsfxsize=7cm 
\epsfbox{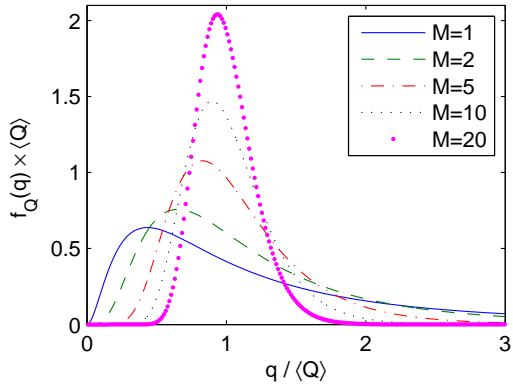}\ \\
\end{tabular}
\end{center}
{
\caption{\label{fig:PDFQ_unstirred} \small
Scaled PDF $f_Q(q)$ {of} normalized $Q$ (i.e., in units $\langle Q \rangle = [3M/(2M-1)][(hV/({\mu_{\rmw,\rmr} \delta_\rmw S})]$) for selected values of $M$.}}
\end{figure}

Regarding the relationship between $U$ and $P_\rmd$, the boundary field $\ul{H}_t(\ul{r}_{S})$ is deterministically related to 
the interior $\ul{H}(\ul{r})$, because of field continuity and EM boundary conditions. 
Therefore, $U$ and $P_{\rm d}$ are {\em not\/} statistically independent. 
Nevertheless, their joint PDF can always be expressed as the product of the marginal PDF of $U$ and the conditional PDF of $P_\rmd$ given $U$, viz., $f_{U,P_\rmd}(u,p_\rmd)=f_U(u)f_{P_\rmd|U}(u,p_\rmd|u)$.
Based on this factorization, $f_Q(q)$ is derived in Appendix \ref{app:F} as the ratio of a $\chi^2_{2r}$ distributed  $\omega U$ and a $\chi^2_{2s}$ distributed $P_\rmd$, for general values of $r$ and $s$, resulting in a Fisher--Snedecor F-distribution with $(2r,2s)$ DoF (cf. eqs. (\ref{eq:PDFQ_ifv_sigmas}), (\ref{eq:PDFQ_ifv_mus}), (\ref{eq:PDFQ_final}), (\ref{eq:PDFQ_selfsuff_mu}) and (\ref{eq:PDFQ_selfsuff_sigma})).  
Assigning $r=3M$ and $s=2M$, 
the PDF of $Q$ is then
\bea
f_Q(q) =
\frac{\left ( \frac{2M-1}{3M} \langle Q \rangle \right )^{2M}}{\Beta(3M,2M)} \frac{q^{3M-1}}{\left ( q + \frac{2M-1}{3M} \langle Q \rangle \right )^{5M}},
\label{eq:PDFQ_selfsuff_mu_multimod}
\eea
valid for $M>1/2$, representing a F$(6M,4M)$ PDF. 
Figure \ref{fig:PDFQ_unstirred} shows (\ref{eq:PDFQ_selfsuff_mu_multimod}) for selected values of $M$. For $M \sim 1$, significant positive skewness and smaller kurtosis are observed, whereas for $M\rightarrow+\infty$, the PDF evolves to Gaussian normality.  

\begin{figure}[htb] \begin{center} \begin{tabular}{l}
\ \epsfxsize=7cm 
\epsfbox{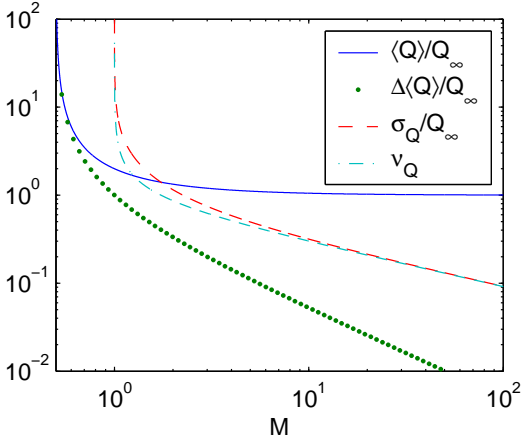}\ \\
\end{tabular}
\end{center}
{
\caption{\label{fig:statsQ} \small
Mean, residual mean, standard deviation and coefficient of variation of $Q$ as a function of $M$, normalized by $Q_\infty$.}}
\end{figure}

The (arithmetic) mean value, standard deviation and coefficient of variation for (\ref{eq:PDFQ_selfsuff_mu_multimod})
follow from (\ref{eq:avgQ})--(\ref{eq:nuQ}) as
\bea
\mu_Q &\equiv& \langle Q \rangle = \frac{3M}{2M-1} ~\frac{h \thinspace V}{\mu_{\rmw,\rmr} \delta_\rmw S},~~~(M>1/2)\label{eq:avgQ_multi}\\
\sigma_Q &=& \sqrt{\frac{3M(3M+1)}{(2M-1)(2M-2)}-\left ( \frac{3M}{2M-1} \right )^2}\frac{h\thinspace V}{\mu_{\rmw,\rmr} \delta_\rmw S},\nonumber\\&~&~~~~~~~~~~~~~~~~~~~~~~~~~~~~~~~~~~~~~~~~~(M>1) \label{eq:stdQ_multi}\\
\nu_Q &=& \sqrt{\frac{(2M-1)(3M+1)}{3M(2M-2)}-1},~~~(M>1).
\label{eq:nuQ_multi}
\eea
Their corresponding limit expressions for $M \gg 1$ are
\bea
\mu_Q &\rightarrow& \frac{3}{2} \left ( 1 + \frac{1}{2M} \right ) \frac{h\thinspace V}{\mu_{\rmw,\rmr} \delta_\rmw S}
\label{eq:avgQ_multi_limit}\\
\sigma_Q &\rightarrow& \sqrt{\frac{15}{8M}} ~\frac{h \thinspace V}{\mu_{\rmw,\rmr} \delta_\rmw S}\label{eq:stdQ_multi_limit}\\
\nu_Q &\rightarrow& \sqrt{\frac{5}{6M}}.\label{eq:nuQ_multi_limit}
\eea
The dependencies of (\ref{eq:avgQ_multi})--{(\ref{eq:nuQ_multi_limit})} on $M$ are shown in Fig. \ref{fig:statsQ}.
For $M \rightarrow +\infty$, the mean $\langle Q \rangle$ reduces asymptotically to half its value for $M=1$, i.e., to 
\bea
Q_\infty \stackrel{\Delta}{=} \frac{3 \thinspace h \thinspace V}{2 \mu_{\rmw,\rmr} \delta_\rmw S}.
 \label{eq:avgQ_asymp}
\eea
The residual mean $\Delta \langle Q \rangle /Q_\infty \stackrel{\Delta}{=} (\langle Q \rangle - Q_\infty)/Q_\infty = 1/(2M-1)$ is positive and asymptotically inversely proportional to $M$.
For $h=1$, the result (\ref{eq:avgQ_asymp}) was previously obtained \cite{lamb1}, \cite{liu1}, \cite{arnaQ}, \cite{hill1998}, whereas the finding that $\langle Q \rangle = 2\thinspace Q_\infty$ when $M=1$ agrees with the findings in \cite{boos1}. This demonstrates that
consistent asymptotic results are retrieved. In Sec. \ref{sec:centrality}, $\langle Q \rangle$ will be compared to other measures of centrality for $Q$. 
For $M=1$, the $\sigma_Q$ and $\nu_Q$ are undefined ($\sigma_Q, \nu_Q \rightarrow +\infty$), whereas for $M\rightarrow +\infty$ they asymptotically approach zero proportionally to $1/\sqrt{M}$, i.e., more slowly than $\Delta \langle Q \rangle /Q_\infty$.
In summary, for $M\rightarrow+\infty$, the limit PDF of $Q$ is a normal distribution ${\cal N} ( \mu_Q, \sigma_Q)$, i.e.,
\bea
f_Q(q) \sim {\cal N} \left ( \left ( 1 + \frac{1}{2M} \right ) Q_\infty,~ \sqrt{\frac{5}{6M}} Q_\infty \right ).
\label{eq:PDF_asymp}
\eea

Parenthetically, if $M\not \gg 1$ then the ratio of the averages $\langle \omega U \rangle$ and $\langle P_\rmd | U \rangle$ is substantially different from the averaged ratio $\langle \omega U/ (P_\rmd | U) \rangle$. 
Indeed, for $M=1$ with (\ref{eq:Qsinglemodeunstirred}), (\ref{eq:avgQ_multi}) and (\ref{eq:sigmaX}), i.e., $\langle |H_0|^2 \rangle = 6 \sigma^2_{H^{\prime(\prime)}_\alpha}$ for a $\chi^2_{6}$ distributed $|H_0|^2$ while $\langle |H_{t,0}|^2 \rangle = 4 \sigma^2_{H^{\prime(\prime)}_\alpha}$ for a $\chi^2_{4}$ distributed $|H_{t,0}|^2$, we arrive at 
\bea
\frac{\langle \omega U \rangle}{\langle P_\rmd | U \rangle } &=& \frac{h \thinspace V}{\mu_{\rmw,\rmr} \delta_\rmw S} \frac{\langle |H_0|^2 \rangle}{\langle |H_{t,0}|^2 \rangle} =\frac{3 \thinspace h \thinspace V}{2\mu_{\rmw,\rmr} \delta_\rmw S} \label{eq:tempratioavg}\\
&=& \frac{1}{2}\left \langle \frac{\omega U}{P_\rmd|U} \right \rangle,~~~(M=1).
\label{eq:Qsinglemodestirred}
\eea
In view of (\ref{eq:avgQ_asymp}), this result shows that replacing $\langle \omega U/ (P_\rmd|U) \rangle$ by $\langle \omega U \rangle/\langle P_\rmd | U \rangle$ as in (\ref{eq:defQeff}) is an approximation, but justifiable when $M\gg 1$, e.g., in overmoded regime. 
In fact, comparing (\ref{eq:defQeff}), (\ref{eq:avgQ_asymp}) and {(\ref{eq:tempratioavg})} using (\ref{eq:sigmaX}) for general $M$ shows that $Q_{\rm eff} \equiv Q_\infty$ for {\em any\/} $M$. Thus, the definition of composite $Q_{\rm eff}$ neglects the effect of $M$ on the fluctuation and value of $Q$. 

\begin{figure}[htb] \begin{center} \begin{tabular}{c}
\ \epsfxsize=7.5cm 
\epsfbox{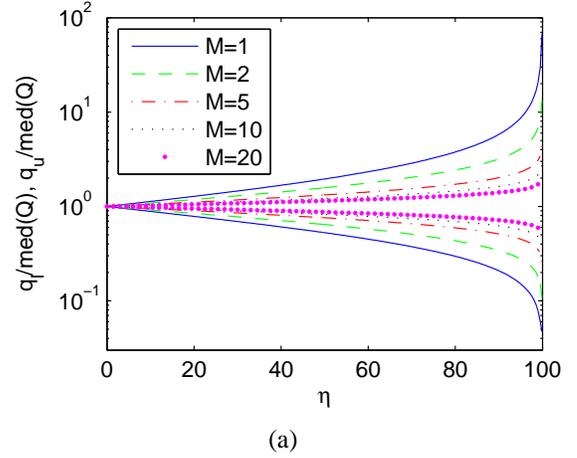}\ \\
(a)\\
\ \epsfxsize=7.5cm 
\epsfbox{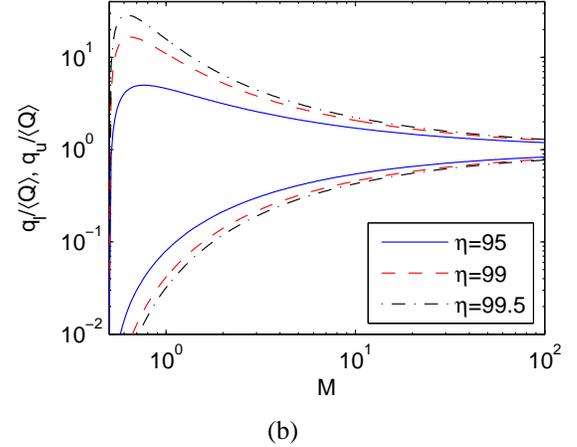}\ \\
(b)
\end{tabular}
\end{center}
{
\caption{\label{fig:confintQ_unstirred} 
\small
Upper ($q_\rmu$) and lower ($q_\ell$) boundaries of $\eta\%$-confidence intervals of $Q$: 
(a) normalized by $\med(Q)$, as a function of confidence level $\eta\%$, for selected values of $M$; 
(b) normalized by $\langle Q \rangle$, as a function of $M$, for $95\%$, $99\%$ and $99.5\%$ confidence levels.}
}
\end{figure}

As an alternative to $\sigma_Q$, the spread of $Q$ can also be expressed by an $\eta\%$-confidence interval {for} $Q$. For a chosen confidence level $\eta$, the boundaries $q_{\ell}$ and $q_{u}$ of this interval are calculated by inverting the cumulative distribution function (CDF) (\ref{eq:CDF_selfsuff_mu})--(\ref{eq:defxi}), i.e., by numerically solving 
\bea
F_Q(q_{\ell,u}) \equiv 1 - I_{\xi_{\ell,u}}(2M,3M) = \frac{1\pm (\eta/100)}{2}, \label{eq:confint_generalM}
\eea
where 
$
I_{\xi_{\ell,u}}(\cdot,\cdot)
$
is a regularized incomplete beta function with
\bea
\xi_{\ell,u} = \left ( 1 + \frac{3M}{2M-1} \frac{q_{\ell,u}}{\langle Q \rangle} \right )^{-1}.
\label{eq:xi_generalM}
\eea
Figures \ref{fig:confintQ_unstirred}a and \ref{fig:confintQ_unstirred}b show these boundaries normalized by the median of $Q$ (cf. (\ref{eq:medQ})--(\ref{eq:medQ_xi})) or {by} $\langle Q \rangle$,  as a function of $\eta$ or $M$, respectively. The interval width rapidly increases with $\eta$ {most prominently} when $M\sim 1$ and $\eta > 90$. For larger $M$, the spread is considerably reduced because of the effect of aggregation of modes, which can also be achieved through multiple stirring mechanisms, multiple sources (antennas), increased EM losses, etc. 
For $M=1$, $3$, $10$ and $100$, the $95\%$-confidence intervals for $Q/\langle Q \rangle$ are $[0.080,4.599]$, $[0.301,2.590]$, $[0.545,1.713]$ and $[0.833,1.192]$, respectively.

The standard deviations of the complex-valued analytic EM fields $\ul{E}=\ul{E}^\prime-\rmj \ul{E}^{\prime\prime}$ and $\ul{H}=\ul{H}^\prime-\rmj \ul{H}^{\prime\prime}$ can be estimated on the premise that, in steady state, the dissipated power equals the transmitted power, i.e., $P_\rmd(\tau) = P_{\rm Tx} (\tau)$, due to conservation of energy. With (\ref{eq:Pd}) and (\ref{eq:sigmaX}) for $n\equiv s = 2M$, it follows that $\langle P_\rmd|U\rangle = M S \sigma^2_{H^{\prime(\prime)}_\alpha} / 2$, whence
\bea
\sigma^2_{E^{\prime(\prime)}_\alpha} = \frac{2 \sigma_\rmw \delta_\rmw \eta^2_0}{M~S} \langle P_{\rm Tx} \rangle,~~~~
\sigma^2_{H^{\prime(\prime)}_\alpha} =
\frac{2 \sigma_\rmw \delta_\rmw}{M~S} \langle P_{\rm Tx} \rangle,
\label{eq:sigmaHsigmaE}
\eea
where $\eta_0 \stackrel{\Delta}{=} \sqrt{\mu_0/\epsilon_0}$ is the stir averaged\footnote{More accurate estimates for $\sigma^2_{E^{\prime(\prime)}_\alpha}$ and $\sigma^2_{H^{\prime(\prime)}_\alpha}$ are obtained by incorporating the random fluctuations of the input impedance dyadic $\ul{\ul{Z}}(\tau)$ \cite{water1963}--\cite{zheng2006}.} input impedance of the MT/MSRC. Thus, like for $Q$, the standard deviation of the stirred EM field decreases proportionally to $1/\sqrt{M}$. This is a result of intrinsic averaging of fields caused by the simultaneous excitation of modes. 
For the total (3-D vector) fields, $\sigma^2_{E^{\prime(\prime)}} = 3 \sigma^2_{E^{\prime(\prime)}_\alpha}$ and $\sigma^2_{H^{\prime(\prime)}} = 3 \sigma^2_{H^{\prime(\prime)}_\alpha}$.

\subsection{Other measures of location for $Q$\label{sec:centrality}}
Because of the primary practical interest in the central value of $Q$, we explore a few other measures of location (centrality) as alternatives to the arithmetic mean $\langle Q \rangle$.
Compared to such other metrics, $\langle Q \rangle$ represents the `centre of mass' of the PDF and minimizes the expected mean squared deviation of the sample values of $Q$. The $\langle Q \rangle$ is known to provide the most stable measure of centrality when comparing values obtained from different {\em sample\/} sets of data. 
However, it is not the optimal measure of centrality for {\em ensemble\/} data, particularly when the PDF is significantly skewed, as in the case of relatively small $M$. 
In the latter case, the mode (for unimodal data) and the median are more representative parameters.

\begin{figure}[htb] \begin{center} \begin{tabular}{l}
\ \epsfxsize=7.5cm 
\epsfbox{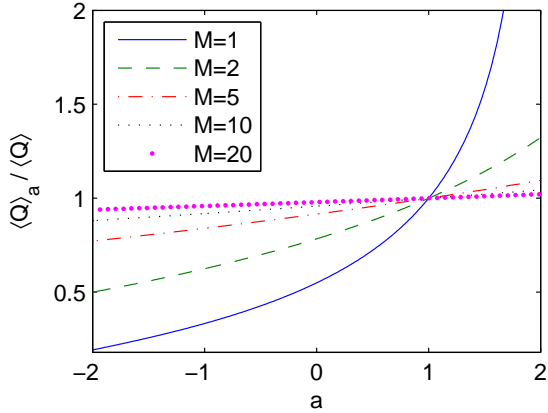}\ \\
\end{tabular}
\end{center}
{
\caption{\label{fig:meansQ} \small
Generalized mean $\langle{Q}\rangle_a$ normalized by arithmetic mean $\langle{Q}\rangle\equiv\langle{Q}\rangle_1$ as a function of $a$, for selected values of $M$.}}
\end{figure}

The generalized mean $\langle Q \rangle_a 
\stackrel{\Delta}{=} \left ( \int^{\infty}_0 q^a f_Q(q) \rmd q \right )^{1/a}$
can be calculated with the aid of (\ref{eq:avgmomQdef}) and (\ref{eq:avgQ_asymp}) as
\bea
\langle Q \rangle_a 
= \frac{2}{3}\left ( \frac{\Gamma(3M+a)\Gamma(2M-a)}{\Gamma(3M)\Gamma(2M)} \right )^{1/a} Q_\infty,
\label{eq:avgmomQ}
\eea
where $a$ is a chosen real parameter. The particular cases $a=-1$, $0$, $1$ and $2$ correspond to the harmonic, geometric, arithmetic and RMS averages, respectively.
Figure \ref{fig:meansQ} shows $\langle{Q}\rangle_a$ as a function of $a$ for selected values of $M$, after normalization with respect to $\langle{Q}\rangle \equiv \langle{Q}\rangle_1$. For any $a<1$, $\langle{Q}\rangle_a$ is always smaller than $\langle{Q}\rangle$, {\it a fortiori} for small $M$. 
For $M\rightarrow +\infty$, all $\langle Q \rangle_a$ merge to $\langle Q \rangle$ irrespective of $a$.

The statistical mode is the most probable (or most frequent) value among the values of the population (or sample data set), i.e., for $\{Q(\tau)\}$ across one rotation of a mode stirrer. Solving $\rmd [f_Q(q)]/ \rmd q = 0$ using (\ref{eq:PDFQ_selfsuff_mu_multimod}), the mode-to-mean ratio is 
\bea
\frac{\mod(Q)}{\langle Q \rangle}= \left ( 1-\frac{1}{3M} \right ) \left ( 1-\frac{2}{2M+1} \right )\rightarrow 1-\frac{4}{3M}.
\label{eq:modQ}
\eea

Another measure of centrality is the median, defined as
$
\med(Q) \stackrel{\Delta}{=} F^{-1}_Q(0.5)
$ and obtained from (\ref{eq:CDF_selfsuff_mu})--(\ref{eq:defxi}) by solving
\bea
I_{\xi_m}(2M, 3M) = \frac{1}{2},
\label{eq:medQ}
\eea
where
\bea
\xi_m \stackrel{\Delta}{=} \left ( 1 + \frac{3M}{2M-1} \frac{\med(Q)}{\langle Q \rangle}\right )^{-1}
.
\label{eq:medQ_xi}
\eea
A numerical approximation of $\med(Q)$ is obtained from (\ref{eq:avgmomQ}) for $a=-0.065$.
Unlike $\langle Q \rangle$, the median minimizes the expected {\em absolute\/} deviation. It is a robust measure of centrality, being less sensitive to the shape of $f_Q(q)$. This is particularly attractive because of the cited difficulties of characterizing the precise PDFs of $U$ and $P_\rmd$ when $M>1$. 

Comparing these metrics, for any $M > 1/2$, the ordering
\bea
\mod(Q) \leq Q_\infty \leq \med(Q) \leq \langle Q \rangle \leq \mathit{Q}_{\rm RMS}
\label{ineq:central}
\eea
applies, together with 
$\mod(Q) < \langle Q \rangle_a < Q_\infty$          when $-1<a<-0.185$,
    $Q_\infty < \langle Q \rangle_a < \med(Q)$      when $-0.185<a<-0.065$, and
$\med(Q) < \langle Q \rangle_a < \langle Q \rangle$ when $-0.065<a<1$.

\begin{figure}[htb] \begin{center} \begin{tabular}{l}
\ \epsfxsize=7.5cm 
\epsfbox{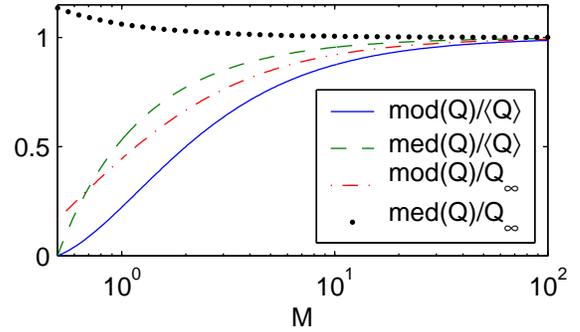}\ \\
\end{tabular}
\end{center}
{
\caption{\label{fig:modtomeanQ_medtomeanQ} \small
Mode-to-mean and median-to-mean ratios $\mod(Q)/\langle{Q}\rangle$, $\med(Q)/\langle{Q}\rangle$, $\mod(Q)/Q_\infty$ and $\med(Q)/Q_\infty$ as functions of $M$.}}
\end{figure}

Figure \ref{fig:modtomeanQ_medtomeanQ} shows that the ratio 
$\mod(Q) / \langle Q \rangle$ increases from $0$ when $M\rightarrow 1/2$, over $2/9$ at $M=1$, to $1$ when $M\rightarrow +\infty$. 
With the same marker values of $M$, this corresponds to $\mod(Q) / Q_\infty$ increasing from $1/6$ over $4/9$ to $1$.
The ratio $\med(Q) / \langle Q \rangle$ increases from $0$ over 
$0.5308$ to $1$, whereas $\med(Q) / Q_\infty$ decreases from $1.135$ over $1.062$ to $1$. 

In practice, experimentally determined values of $Q$ are nearly always reported to be considerably {\em smaller\/} than $Q_{\rm eff} \equiv Q_{\infty}$, typically by a factor $0.2$ or $0.5$ to $1$ \cite{warn1}, \cite{beck1}, \cite{frey1}. On account of (\ref{ineq:central}), choosing $\mod(Q)$ or {\it a fortiori} $\langle Q \rangle_{(a< -2)}$ as an {\it a priori} theoretical estimate may offer better {\it ad hoc} quantitative agreement than $\langle Q \rangle$ and $\med(Q)$. Physically, however, additional loss mechanisms are at the root of lowering $Q$ \cite{lamb1}. 

\subsection{$Q$ in mode-stirred vs. mode-tuned chambers\label{sec:MSvsMT}}
The previous analysis assumed quasi-static operation, through sufficiently slow stepping or scanning (mechanical or electronic).
General considerations of mode-stirred vs. mode-tuned operation focusing on acquisition time and nonstationary effects were given in \cite{iec}, \cite{arnalocavg}, \cite{tich1974}--\cite{arnaJPA}. Here, we comment on aspects of mode stirring relating to $Q$ and its PDF, and we restrict ourselves merely to some general remarks. 

In quasi-stationary MT/MSRCs, the rate of change of the cavity field between stir states is small compared to the rate of energy fill and dissipation, whence the values of $U(\ul{r},\tau)$ and $P_\rmd(\ul{r}_{S},\tau)$ remain unaffected by this rate. 
Quasi-stationary mode stirring may results in purely local uniform temporal averaging of the fields across an interval of stir states $[0,{\cal T}]$ varying with time. 
This averaging does obviously not affect the spatial averagings of $U(\ul{r}|\tau)$ and {$P_\rmd(\ul{r}_{S}|\tau)$}, yielding again (\ref{eq:U_ifv_tau}) and (\ref{eq:Pd_ifv_tau}). 
Ensemble averaging of $U(\tau)$ and $P_\rmd(\tau)$ across $[0,{\cal T}]$ does not affect $\langle U (\tau) \rangle$ and $\langle P_\rmd (\tau) | U (\tau) \rangle$, whereas both $\sigma^2_{U(\tau)}$ and $\sigma^2_{P_\rmd(\tau)|U(\tau)}$ scale by the {\em same\/} variance function $\gamma_{\cal T}$ in quasi-stationary conditions \cite{arnalocavg}.
Consequently, the ratio $\sigma_U/\sigma_{P_\rmd|U}$ and hence $f_Q(q)$ remain unaffected by local averaging. 
However, the existing coupling between TE and TM modes caused by ohmic losses is further increased by continuous rotation, albeit as a second-order effect of velocity of rotation \cite{yild1966}, \cite{noji2004}. In turn, this has a positive but second-order effect on $P_\rmd$. Recent measurements of $\langle Q \rangle$ in an overmoded MSRC \cite{raja1} appear to support these findings, including a marginal but systematic decrease of $\langle Q \rangle$ observed for increased stir speed using the mean power approach.

For nonstationary MT/MSRCs, the situation is more intricate. 
Expanding wave fronts emanating from the source impinge and get partially absorbed by the cavity walls, after traversing the cavity interior. Therefore, the cavity fill and dissipation processes per stir state are not synchronized and may be affected differently during nonstationary stirring in a resonant environment, as boundary and excitation conditions vary in a rapid manner. This situation is to be avoided, in view of the definition (\ref{eq:defQ}) requiring matching (i.e., comparable) pairs of $U(\tau)$ with $P_\rmd(\tau)$. 
Hence, for a nonstationary MSRC, one may need recourse to the physically less meaningful definition (\ref{eq:defQeff}), thereby using nonuniform weighting \cite{arnaPRE2010}. 

\section{Conclusion}
In this paper, we derived a PDF for the $Q$ of a MT/MSRC, with the number of simultaneously excited cavity modes ($M$) as a distribution parameter.
The analysis assumed ideal $\chi^2$ ensemble distributions for stored energy and dissipated power, associated with unbiased circular Gaussian distribution of the stirred cavity field, in space and time (stir domain). Perfectly efficient mode stirring, i.e., a theoretically unlimited number of statistically independent stir states of the {local} field ($N\rightarrow +\infty$), was also assumed but can be relaxed by using sampling distributions to replace the ensemble distributions. 
In practice, this condition on $N$ is usually more closely achieved at short wavelengths ($\lambda \ll V^{1/3}$). Also, strong positive correlation typically exists between $M$ on $N$ in practical stirring techniques, which may leave the expressions for $f_Q(q)$ to be somewhat approximate in practice when $M\sim 1$.

With these idealizations, $Q$ was found to exhibit a Fisher--Snedecor F$(6M,4M)$ PDF, given by (\ref{eq:PDFQ_selfsuff_mu_multimod}).
For general $M$, its arithmetic mean value $\langle Q \rangle$, standard deviation and coefficient of variation were obtained in (\ref{eq:avgQ_multi})--(\ref{eq:nuQ_multi}). For $M\rightarrow +\infty$, the PDF approaches a Gauss normal limit PDF (\ref{eq:PDF_asymp}). 
Confidence intervals for $Q$ were calculated numerically from (\ref{eq:confint_generalM})--(\ref{eq:xi_generalM}) and indicate that the spread of $Q$ can be considerable, even when $M \gg 1$. 
Expressions for the standard deviations of the EM fields were obtained in (\ref{eq:sigmaHsigmaE}). 
Alternative measures of centrality (viz., the generalized mean (\ref{eq:avgmomQ}), mode (\ref{eq:modQ}) and median (\ref{eq:medQ})--(\ref{eq:medQ_xi})) rank according to (\ref{ineq:central}), producing values that are always smaller than $\langle Q \rangle$, for any $M$ but especially for $M\sim 1$. Specifically, the most probable and most robust central values of $Q$ are smaller than $\langle Q \rangle$ by factors ranging from $2/9$ and $0.5308$ (for $M=1$) to $1$ (for $M\rightarrow +\infty$), respectively. 
Compared to the mean and especially the median, the mode and generalized mean with index $a< -2$ agree quantitatively with typical measured values of $Q$ that are much lower than {\it a priori} estimated $\langle Q \rangle$. 

Regarding extensions of the present analysis, the nonorthogonality of the magnetic eigenvectors on the boundary in case of multi-mode excitation is known to affect the total dissipated power. This requires characterization and quantification in the context of dynamic cavities. 
Generalization to nonzero excitation bandwidths is required to quantify the resulting increase of the uncertainty of $Q$. 
The effect of the nonintegrability (`complexity') of real MT/MSRC enclosures -- including the geometry of a mode stirrer -- on $h$ in (\ref{eq:defh}) and the effect of elongation or flattening of the cavity shape on $f_Q(q)$ deserve further attention.
A rigorous generalization to fluctuating $M$ during stirring (cf. Sec. \ref{sec:singlevsmulti}), caused by entrance and exit of modes in the cavity bandwidth \cite{arnaTEMC2003a}, is of interest. The PDF $f_M(m)$ is expected to depend on the eigenfrequency spacing statistics \cite{lyon1969} and their dynamics, which, in turn, depend on the integrability of the cavity shape. The additional uncertainty caused by fluctuating $M$ is expected to increase the width of the confidence interval of $Q$. 
Finally, the (auto)correlation function of $\rho_Q(\Delta\tau)$ provides insight into the {\em rate\/} of fluctuation $\rmd Q/\rmd \tau$, i.e., the stir sensitivity of $Q(\tau)$.

\appendix
\section{Self-sufficient forms of F-distributions for $Q$ based on idealized cavity field distributions\label{app:F}}
The PDF of $Q\stackrel{\Delta}{=}\omega U/P_\rmd$ can be calculated from the joint PDF\footnote{It is assumed here that the excitation is sufficiently narrowband so that $\omega$ can be considered to be a deterministic (constant) quantity.} 
$f_{\omega U,P_\rmd}( \omega u, p_\rmd )$ of $\omega U$ and $P_\rmd \equiv \omega U/Q$, as
\bea
f_Q(q) = \frac{1}{q^2} \int^{+\infty}_0 f_{\omega U,P_\rmd}\left ( \omega u, \frac{\omega u}{q} \right ) |\omega u| ~ \rmd (\omega u),
\label{eq:defPDFQ}
\eea
in which
$f_{\omega U,P_\rmd}(\omega u,p_\rmd) = f_{\omega U}(\omega u) f_{P_\rmd|\omega U}(p_\rmd|\omega u)$, with 
$f_{Y} (y) = f_U(u=y/\omega)/|\omega|$ for the auxiliary variate $Y\stackrel{\Delta}{=} \omega U$ and 
$f_{P_\rmd|\omega U}(p_\rmd|\omega u) = $ $f_{P_\rmd|U}(p_\rmd|u)$ for deterministic $\omega$.
For ideal Gaussian interior and surface fields, the associated energy density $U$ and conditional dissipated power $P_{\rm d}|U$ exhibit $\chi^2_{2r}$ and $\chi^2_{2s}$ PDFs, respectively, i.e., in self-sufficient form \cite{arnalocavg}
\bea
f_U(u) &=& \frac{r^{r/2}}{\Gamma(r) \sigma^r_U} u^{r-1} \exp \left ( - \frac{\sqrt{r}}{\sigma_U}\thinspace u \right )\label{eq:PDFU}\\
f_{P_\rmd|U}(p_\rmd|u) &=& \frac{s^{s/2}}{\Gamma(s) \sigma^s_{P_\rmd|U}} {p_\rmd}^{s-1} \exp \left ( - \frac{\sqrt{s}}{\sigma_{P_\rmd|U}} \thinspace p_\rmd \right ),~~~~\label{eq:PDFPd}
\eea
with which (\ref{eq:defPDFQ}) reduces, with the aid of \cite[(3.351.3)]{grad1}, to
\bea
f_Q(q)  &=& \frac{\omega}{q^2} \int^{+\infty}_0 f_{U} ( u ) f_{P_\rmd|U}\left ( \frac{\omega u}{q} \right ) u ~ \rmd u\\
&=& 
\frac{1}{\Beta(r,s)} \left ( \sqrt{\frac{s}{r}} \frac{\omega \sigma_U}{\sigma_{P_\rmd|U}}\right )^s \frac{q^{r-1}}{\left ( q + \sqrt{\frac{s}{r}} \frac{\omega \sigma_U}{\sigma_{P_\rmd|U}} \right )^{r+s}},
\nonumber\\
\label{eq:PDFQ_ifv_sigmas}
\eea
where the complete beta function is calculated as
\bea
\Beta(r,s) = \frac{\Gamma(s) \Gamma(r)}{\Gamma(s+r)}  = \frac{(s-1)! ~ (r-1)!}{(s+r-1)!}.
\eea
Alternatively, (\ref{eq:PDFQ_ifv_sigmas}) can be expressed in terms of average values serving as distribution parameters, because any $\chi^2_{2n}$ distributed variate $X$ -- being the sum of squares of $2n$ independent and identically distributed ${\cal N}(0,\sigma)$ variates -- has its standard deviation and mean value related via \cite{arnaTQE2}
\bea
\sigma_X = \frac{\langle X \rangle}{\sqrt{n}} = 2 \sqrt{n} \thinspace \sigma^2, 
\label{eq:sigmaX}
\eea
as a result of which (\ref{eq:PDFQ_ifv_sigmas}) can be written as
\bea
f_Q(q) =
\frac{1}{\Beta(r,s)} \left ( \frac{s}{r} \frac{\omega \langle U \rangle}{\langle{P_\rmd|U}\rangle}\right )^s \frac{q^{r-1}}{\left ( q + {\frac{s}{r}} \frac{\omega \langle U \rangle}{\langle {P_\rmd|U} \rangle} \right )^{r+s}}
.
\label{eq:PDFQ_ifv_mus}
\eea
Equations (\ref{eq:PDFQ_ifv_sigmas}) and (\ref{eq:PDFQ_ifv_mus}) represent a Fisher--Snedecor F-distribution with $(2r,2s)$ degrees of freedom.

Now we wish to express {the ratio of} the field-dependent distribution parameters $\sigma_U/\sigma_{P_\rmd|U}$ or $\langle U \rangle / \langle{P_\rmd|U}\rangle$ in (\ref{eq:PDFQ_ifv_sigmas}) or (\ref{eq:PDFQ_ifv_mus}) in terms of field-independent cavity design parameters. The evaluation in Appendix \ref{app:omegasigamUoverPd} yields these ratios as (\ref{eq:temp11}).
Hence, (\ref{eq:PDFQ_ifv_sigmas}) and (\ref{eq:PDFQ_ifv_mus}) can be written as
\bea
f_Q(q) =
\frac{1}{\Beta(r,s)} \left ( \frac{h \thinspace V}{\mu_{\rmw,\rmr} \delta_\rmw S}\right )^s \frac{q^{r-1}}{\left ( q + \frac{h \thinspace V}{\mu_{\rmw,\rmr} \delta_\rmw S} \right )^{r+s}}
,
\label{eq:PDFQ_final}
\eea
in which $h$ is the average shape factor of the stirred cavity, defined by (\ref{eq:defh}).

With the moments of $Q$ of order $i$ defined as $\langle Q^i \rangle \stackrel{\Delta}{=} \int^{\infty}_0 q^i f_Q(q) {\rm d}q$ and evaluated using \cite[(3.194.3)]{grad1} as
\bea
\langle Q^i \rangle =  \frac{\Gamma(r+i)\Gamma(s-i)}{\Gamma(r)\Gamma(s)} \left ( \frac{h \thinspace V}{\mu_{\rmw,\rmr} \delta_\rmw S} \right )^i,
\label{eq:avgmomQdef}
\eea
the arithmetic mean, the standard deviation and the coefficient of variation are obtained as
\bea 
\mu_Q &\equiv& \langle Q \rangle  =
\frac{r}{s-1} ~ \frac{h \thinspace V}{\mu_{\rmw,\rmr} \delta_\rmw S},~~~(s>1)\label{eq:avgQ}\\
\sigma_Q &=& \sqrt{\langle Q^2 \rangle - \langle Q \rangle^2} \nonumber\\
&=& \sqrt{\frac{r(r+1)}{(s-1)(s-2)}-\left ( \frac{r}{s-1} \right )^2}\frac{h \thinspace V}{\mu_{\rmw,\rmr} \delta_\rmw S},\nonumber\\&~&~~~~~~~~~~~~~~~~~~~~~~~~~~~~~~~~~~~~~~(s>2) \label{eq:stdQ}\\
\nu_Q &=& \frac{\sigma_Q}{\mu_Q} = \sqrt{\frac{(s-1)(r+1)}{(s-2)r}-1},~~~(s>2)\label{eq:nuQ}
\eea
respectively. The form (\ref{eq:PDFQ_final}) can therefore be rewritten in self-sufficient format in terms of $\langle Q \rangle$, i.e.,
\bea
f_Q(q) =
\frac{1}{\Beta(r,s)} \left ( \frac{s-1}{r} \langle Q \rangle \right )^s \frac{q^{r-1}}{\left ( q + \frac{s-1}{r} \langle Q \rangle \right )^{r+s}}
\label{eq:PDFQ_selfsuff_mu}
\eea
valid for $s>1$, or alternatively in terms of $\sigma_Q$, as
\bea
f_Q(q) &=& 
\frac{1}{\Beta(r,s)} \left ( \sqrt{\frac{(s-1)^2(s-2) }{r(r+1)(s-1)-r^2(s-2)}} \sigma_Q \right )^{s} \nonumber\\
&~&\times
\frac{q^{r-1}}{\left ( q +  \sqrt{\frac{(s-1)^2(s-2) }{r(r+1)(s-1)-r^2(s-2)}} \sigma_Q \right )^{r+s}},
\label{eq:PDFQ_selfsuff_sigma}
\eea
valid for $s>2$.
Both (\ref{eq:PDFQ_selfsuff_mu}) and (\ref{eq:PDFQ_selfsuff_sigma}) depend neither on field parameters nor on cavity design parameters.

The CDF of $Q$ corresponding to the form (\ref{eq:PDFQ_selfsuff_mu}) is
\bea
F_Q(q) = 1-I_\xi(s,r),~~~~(s>1) \label{eq:CDF_selfsuff_mu}
\eea
where 
\bea
I_\xi(s,r) \stackrel{\Delta}{=} \frac{1}{ \Beta(s,r)} \int^\xi_0 t^{s-1} (1-t)^{r-1} \rmd t
\label{eq:betaincdef}
\eea
is the regularized incomplete beta function with
\bea
\xi \stackrel{\Delta}{=} \left ( 1 + \frac{r}{s-1} \frac{q}{\langle{Q} \rangle} \right )^{-1} 
.
\label{eq:defxi}
\eea

\appendix
\section{Expression of ${\omega \sigma_U}/{\sigma_{P_\rmd|U}}$ in function of cavity design parameters\label{app:omegasigamUoverPd}}
We consider statistically independent, circular Gauss normal, analytical fields $\ul{E}$ and $\ul{H}$.
From (\ref{eq:U}), (\ref{eq:Pd}) and (\ref{eq:amplit}), 
\bea
\frac{\omega\langle U \rangle}{\langle P_\rmd|U \rangle} 
&=& \frac{\omega\frac{\mu_0}{2} \langle \int_V |\ul{H}(\ul{r})|^2 \rmd V \rangle}{\frac{1}{2} \sqrt{\frac{\omega\mu_\rmw}{2\sigma}} \langle \int_S |\ul{H}_t(\ul{r}_S)|^2 \rmd S \rangle }\nonumber\\
&=& 
\sqrt{\frac{2\thinspace\omega \mu_0 \sigma}{\mu_{\rmw,\rmr}}}
\frac{\langle \int_V |\ul{H}(\ul{r})|^2 \rmd V \rangle}{\langle \int_S |\ul{H}_t(\ul{r}_S)|^2 \rmd S \rangle }
.
\label{eq:temp1}
\eea
For general multimode excitation, 
\bea
|\ul{H}(\ul{r})|^2 &=& \left | \sum^{(M)}_{mnp} H_{mnp,0} ~\ul{\phi}_{mnp}(\ul{r}) \right |^2 
\eea
where $\sum^{(M)}_{mnp}$ denotes a sum over $M$ simultaneously excited modes.
Upon ensemble averaging, the cross product terms (incoherent field terms) in the summation vanish and we obtain
\bea
\left \langle \int_V |\ul{H}(\ul{r})|^2 \rmd V \right \rangle
&=& \left \langle \int_V  \sum^{(M)}_{mnp} |{H}_{mnp,0}|^2 |\ul{\phi}_{mnp}(\ul{r})|^2 \rmd V \right \rangle\nonumber\\
&\simeq& \left \langle |{H}_{0}|^2 \int_V \sum^{(M)}_{mnp} |\ul{\phi}_{mnp}(\ul{r})|^2 \rmd V \right \rangle
\label{eq:temp}\\
&=& M \left \langle |{H}_{0}|^2 \right \rangle \left \langle \int_V |\ul{\phi}(\ul{r})|^2 \rmd V \right \rangle
.
\label{eq:temp2}
\label{eq:avgH}
\eea
The approximation in (\ref{eq:temp}) arises as a result of replacing the superposition of modal amplitudes $H_{mnp,0}$ by 
a single random amplitude $H_0$, for arbitrary $\tau$.
The equality (\ref{eq:temp2}) holds provided that the output impedance of the source is matched to the input impedance of the cavity for each stir state $\tau$, such that each source amplitude $H_{mnp,0}$ is independent of $\tau$ and, hence, independent of the variation of $\ul{\phi}_{mnp}$ as a function of $\tau$. The modal triplets can be omitted when the integral of $|\ul{\phi}_{mnp}|^2$ is independent of $k_{mnp}$ (cf. Sec. \ref{sec:unstirred}).
Similarly,
\bea
\left \langle \int_S |\ul{H}(\ul{r}_S)|^2 \rmd S \right \rangle
= M \left \langle |{H}_{t,0}|^2 \right \rangle \left \langle \int_S |\ul{\phi}(\ul{r}_S)|^2 \rmd S \right \rangle
.
\label{eq:avgHS}
\eea
Substituting (\ref{eq:avgH}), (\ref{eq:avgHS}) and (\ref{eq:sigmaX}) into (\ref{eq:temp1}) yields 
\bea
\frac{\omega \sigma_U}{\sigma_{P_\rmd|U}} &=& \sqrt{\frac{s}{r}} \frac{\omega \langle U \rangle}{\langle{P_\rmd|U} \rangle}
= 
\sqrt{\frac{r}{s}} \frac{h\thinspace V}{\mu_{\rmw,\rmr} \delta_\rmw S} \label{eq:temp11}\\
&=& 
\sqrt{\frac{r}{s}} \frac{2}{\mu_{\rmw,\rmr} \delta_\rmw} \frac{\langle \int_V |\ul{\phi}(\ul{r})|^2 \rmd V \rangle}{\langle \int_S |\ul{\phi}(\ul{r}_S)|^2 \rmd S \rangle}
,
\label{eq:temp10}
\eea
where 
\bea
h
\stackrel{\Delta}{=} \frac{2~\langle \langle |\ul{\phi}(\ul{r})|^2 \rangle_V \rangle}{\langle \langle |\ul{\phi}(\ul{r}_S)|^2 \rangle_S \rangle}
\label{eq:defh}
\eea
defines an average geometry (shape) factor of the stirred cavity that maintains constant $V$ and $S$.
For a wall-stirred rectangular cavity, $|\ul{\phi}(\ul{r})|^2 = \sin^2(k_x x) \sin^2(k_y y) \sin^2(k_z z)$ whence $h=1$ and (\ref{eq:temp11}) becomes
\bea
\frac{\omega \sigma_U}{\sigma_{P_\rmd|U}} = 
\sqrt{\frac{3}{2}} \frac{V}{\mu_{\rmw,\rmr} \delta_\rmw S}
.
\eea

\appendix
\section{Estimation of $M$ \label{app:estM}}
For practical application, an {\it a priori} estimate of the number of simultaneously excited cavity modes, $M$, based on cavity design parameters is of interest.

\begin{figure}[htb] \begin{center} \begin{tabular}{l}
\ \epsfxsize=7.5cm 
\epsfbox{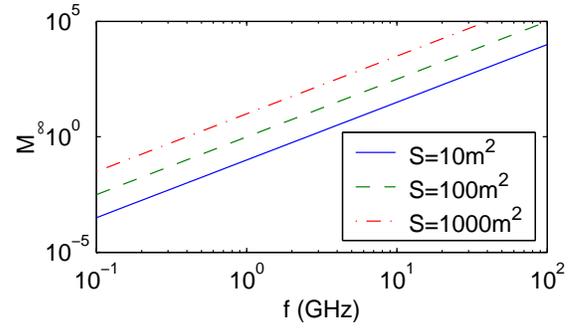}\ \\
\end{tabular}
\end{center}
{
\caption{\label{fig:Mifvf} \small
$M_\infty(f)$ for a rectangular cavity with $\sigma_\rmw=10^6$ S/m and $\mu_{\rmw,\rmr}=1$ at selected values of $S$.}
}
\end{figure}

The average mode density inside a cavity of volume $V$ and surface $S$ operated at CW frequency $f$ is given by the generalized Weyl density \cite{luko1}, \cite{arnaTEMC2001} as
\bea
\frac{{\rm d}{M_\rmc}(f)}{{\rm d}f}
&=&
\frac{8 \pi V}{\rmc} \left ( \frac{f}{\rmc} \right )^2 +
\left [
- \frac{4}{3\pi \rmc} \int\int_{\partial V} \frac{{\rm d}
s}{\overline{\varrho}(\underline{r})} 
\right.
\nonumber\\
&~&
\left. 
+
\frac{1}{6 \pi \rmc} \int_{\partial S}
\frac{[\pi - \varphi(\underline{r})][\pi-5\varphi(\underline{r})] {\rm d}l }{\varphi(\underline{r})}\right ] 
,~~~
\label{eq:modedens}
\eea
up to terms of order $(f/\rmc)^{-2}$ or smaller, where $M_\rmc$ represents the cumulative spectral mode count from dc to $f$.
The other symbols in (\ref{eq:modedens}) were defined in \cite{arnaTEMC2001}.
For $\lambda \ll V^{1/3}$, i.e., $V(f/\rmc)^3 \gg 1$, the number of modes $\delta {M_\rmc}$ inside a narrow band of width $\delta f = \langle f/ Q(f) \rangle \simeq f/\langle Q(f) \rangle$ can be approximated by retaining only the first term in (\ref{eq:modedens}) {and using (\ref{eq:avgQ})} to yield
\bea
\delta {M_\rmc}(f) \simeq \frac{\rmd {M_\rmc}(f)}{\rmd f} \delta f = {8 \pi} \frac{s-1}{r} ~ \frac{\mu_{\rmw,\rmr} \delta_\rmw S}{h} \left ( \frac{f}{\rmc} \right )^3.
\label{eq:deltaMf}
\eea
A CW source excites\footnote{Note that $\delta f$ is defined as the width at half height. Excitation near the edges of the band $\delta f$ is not as strong as near the centre of the band but may contribute through multiplication of $\delta {M_\rmc}$ by a suitable factor of order unity.} 
all overlapping modes within $\delta f$ simultaneously, hence $\delta {M_\rmc}=M$. With $r=3M$ and $s=2M$, (\ref{eq:deltaMf}) yields $M$ as a solution of the quadratic
\bea 
3 M^2 - 2 b M + b = 0,
\eea
where
\bea
b \stackrel{\Delta}{=}  8\pi \frac{\mu_{\rmw,\rmr} \delta_\rmw S}{h} \left ( \frac{f}{\rmc} \right )^3
.
\eea
On physical grounds, the solution with the negative sign is discarded because $M$ is known to increase with $f$. Therefore,
\bea
M(f) = \frac{b + \sqrt{b^2-3b}}{3}
\eea
whose asymptotic expression for $f \rightarrow +\infty$ is $2b/3$, i.e.,
\bea
M_\infty (f) = \frac{16\pi}{3} \frac{\mu_{\rmw,\rmr} \delta_\rmw S}{h} \left ( \frac{f}{\rmc} \right )^3
= \frac{8\pi V} {Q_\infty} \left ( \frac{f}{\rmc} \right )^3,
\eea
i.e., proportional to $f^{5/2}$ {and $\sigma^{-1/2}_\rmw$}. 
Figure \ref{fig:Mifvf} illustrates the rapid increase of $M_\infty$ with $f$, for selected values of $S$ for a rectangular cavity with $\sigma_\rmw=10^6$ S/m and $\mu_{\rmw,\rmr}=1$. For $S=100$ m$^2$, the predicted values of $M_\infty$ at $f=0.1$, $1$ and $10$ GHz are $0.0031$, $0.99$ and $313$, respectively.

Large values of $M$ may occur, even at relatively low modal spectral densities and narrowband excitation, provided the spectral overlap of modes is sufficiently high \cite{arnaTEMC2001} (which depends on ohmic dissipation and leakage of the cavity), or when the excitation bandwidth is large relative to the average spectral modal spacing.
For ultra-low loss enclosures (e.g., superconducting or laser cavities), the modal overlap can be small or nonexistent, even at very high frequencies. Therefore, the value of $M$ need not always be exceedingly large, even in overmoded conditions. For RF and microwave metal cavities, however, the losses and modal overlap are larger whence $M$ is then typically large.
Since the value of $M$ can also be related to the number of coherence cells in $V$, its value can also be estimated by dividing $V$ by the size of such cells \cite{arnaTEMC2002}, \cite{arnaPRE0}.

\end{document}